\documentclass[a4paper,useAMS,usenatbib]{mn2e}
\usepackage[totalwidth=480pt, totalheight=680pt]{geometry}
\usepackage{graphicx}
\usepackage{color}
\usepackage{mathtools}
\def\gr{{$\gamma$-ray}}

\def\nup{$\nu_{\rm peak}^S$}
\newdimen\digitwidth
\setbox0=\hbox{2}
\digitwidth=\wd0
\catcode `#=\active
\def#{\kern\digitwidth}

\newcommand{\fermi}{{\it Fermi}}

\newcommand{\paperone}{{Paper I}}

\newcommand{\paperthree}{{Paper III}}
\newcommand{\paperfour}{{Paper IV}}
\newcommand{\eqb}{\begin{eqnarray}}
\newcommand{\eqe}{\end{eqnarray}}
\newcommand{\en}{E_{\nu}}
\newcommand{\ep}{E_{\nu, p}}
\newcommand{\ns}{\nu_{\rm peak}^S}
\newcommand{\pg}{ p \pi}

\title[A simplified view of blazars: the neutrino background]{A simplified view of
  blazars: the neutrino background}
 \author[P. Padovani et al.]{P. Padovani$^{1,2}$\thanks{E-mail:
ppadovan@eso.org}, M. Petropoulou$^{3}$\thanks{Einstein Postdoctoral Fellow}, P. Giommi$^{4,5}$, 
E. Resconi$^6$\\
$^{1}$European Southern Observatory, Karl-Schwarzschild-Str. 2,
D-85748 Garching bei M\"unchen, Germany\\
$^{2}$Associated to INAF - Osservatorio Astronomico di Roma, via Frascati 33,
I-00040
Monteporzio Catone, Italy\\
$^{3}$Department of Physics and Astronomy, Purdue University, 525 Northwestern
Avenue, West Lafayette, IN 47907, USA\\
$^{4}$ASI Science Data Center, via del Politecnico s.n.c., I-00133 Roma Italy \\
$^{5}$ICRANet-Rio, CBPF, Rua Dr. Xavier Sigaud 150, 22290-180 Rio de Janeiro, Brazil\\
$^6$Technische Universit{\"a}t M{\"u}nchen, Physik-Department, James-Frank-Str. 1, 
D-85748 Garching bei M{\"u}nchen, Germany}

\begin{document}

\date{Accepted ... Received ...; in original form ...}

\pagerange{\pageref{firstpage}--\pageref{lastpage}} \pubyear{2015}

\maketitle

\label{firstpage}

\begin{abstract}
Blazars have been suggested as possible neutrino sources long before the recent
IceCube discovery of high-energy neutrinos. We re-examine this possibility
within a new framework built upon the {\it blazar simplified view} and a
self-consistent modelling of neutrino emission from individual sources. The
former is a recently proposed paradigm that explains the diverse statistical
properties of blazars adopting minimal assumptions on blazars' physical and
geometrical properties. This view, tested through detailed Monte Carlo
simulations, reproduces the main features of radio, X-ray, and $\gamma$-ray
blazar surveys and also the extragalactic $\gamma$-ray background at energies
$\ga 10$ GeV. Here we add a hadronic component for neutrino production and
estimate the neutrino emission from BL Lacs as a class, ``calibrated'' by
fitting the spectral energy distributions of a preselected sample of BL~Lac
objects and their (putative) neutrino spectra. Unlike all previous papers on
this topic, the neutrino background is then derived by summing up at a given
energy the fluxes of each BL Lac in the simulation, all characterised by their
own redshift, synchrotron peak energy, $\gamma$-ray flux, etc. Our main result
is that BL Lacs as a class can explain the neutrino background seen by IceCube
above $\sim 0.5$ PeV while they only contribute $\sim 10\%$ at lower energies,
leaving room to some other population(s)/physical mechanism. However, one cannot
also exclude the possibility that individual BL Lacs still make a contribution
at the $\approx 20\%$ level to the IceCube low-energy events. Our scenario makes
specific predictions testable in the next few years.
\end{abstract}

\section{Introduction}\label{intro}

Blazars are a class of Active Galactic Nuclei (AGN), which host a jet oriented
at a small angle with respect to the line of sight. Highly relativistic
particles moving within the jet and in a magnetic field emit non-thermal
radiation \citep{bla78,UP95}. This is at variance with most other AGN whose
energy is mainly thermal and produced through accretion of matter onto a
supermassive black hole. Because of their peculiar orientation and highly
relativistic state, blazars are characterised by distinctive and extreme
observational properties, including superluminal motion, large and rapid
variability, and strong emission over the entire electromagnetic spectrum. The
two main blazar sub-classes, namely BL Lacertae objects (BL Lacs) and
flat-spectrum radio quasars (FSRQ), differ mostly in their optical spectra,
with the latter displaying strong, broad emission lines and the former instead
being characterised by optical spectra showing at most weak emission lines,
sometimes exhibiting absorption features, and in many cases being completely
featureless.

The spectral energy distributions (SEDs) of blazars are composed of two broad
humps, a low-energy and a high-energy one. The peak of the low-energy hump
(\nup) can occur at widely different frequencies, ranging from about $\sim
10^{12.5}$~Hz ($\sim 0.01$ eV) to $\sim 10^{18.5}$~Hz ($\sim13$ keV).  The
high-energy hump, which may extend up to $\sim 10$ TeV, has a peak energy that
ranges between $\sim 10^{20}$~Hz ($\sim 0.4$ MeV) to $\sim 10^{26}$~Hz ($\sim
0.4$ TeV)\citep{GiommiPlanck,Arsioli2015}. Based on the rest-frame value of
\nup, BL Lacs can be further divided into Low energy peaked (LBL) sources
(\nup~$<10^{14}$~Hz [$<$ 0.4 eV]), Intermediate ($10^{14}$~Hz$<$ \nup~$<
10^{15}$~Hz [0.4 eV $<$ \nup $<$ 4 eV)] and High (\nup~ $> 10^{15}$~Hz [$>$ 4
  eV]) energy peaked (IBL and HBL) sources respectively \citep{padgio95}. It is
generally accepted that the low-energy blazar emission is the result of electron
synchrotron radiation, with the peak frequency reflecting the maximum energy at
which electrons can be accelerated \citep[e.g.][]{GiommiPlanck}. On the other
hand, the origin of their high-energy emission is still under debate. In the
conventional leptonic scenarios, $\gamma$-ray emission is thought to be due to
inverse Compton radiation \citep[e.g.][]{maraschietal92, sikoraetal94}
whereas in leptohadronic scenarios it may be the result of proton synchrotron
radiation \citep[]{aharonian00, mueckeprotheroe01} or may have a photohadronic
origin \citep[e.g.][and references therein]{Petro_2015}.  Given that the
presence of relativistic electrons is necessary to explain at least the
low-energy hump of the SED, it is reasonable to assume that proton acceleration
takes also place in blazar jets (\citealt{biermannstrittmatter87, sironi13,
  globus14} and references, therein).  Although leptohadronic models bear
several attractive features from the theoretical point of view
\citep[e.g.][]{halzen97}, these are not sufficient to rule out leptonic models.
The most promising way to settle the issue of proton acceleration in blazars is
through high-energy neutrino experiments, since neutrino emission is an
unavoidable outcome of leptohadronic models.

\cite{Pad_2014} have recently extended the domain over which blazars could be
relevant astrophysical sources into neutrino territory. Namely, on the basis of a joint
positional and energetic diagnostic, they have suggested a possible association
between eight BL Lacs (all HBL) and seven neutrino events reported by the
IceCube collaboration \citep{ICECube14}. Following up on this idea,
\cite{Petro_2015} have modelled the SEDs of six of these BL Lacs using a one-zone
leptohadronic model and mostly nearly simultaneous data. The neutrino flux for
each BL Lac was self-consistently calculated, using photon and proton
distributions specifically derived for every individual source. The SEDs of the
sources, although different in shape and flux, were all well fitted by the model
using reasonable parameter values. Moreover, the model-predicted neutrino flux
and energy for these sources were of the same order of magnitude as those of the
IceCube neutrinos. In two cases, i.e. MKN 421 and H 1914$-$194, a suggestively
good agreement between the model prediction and the neutrino flux was found.

The hypothesis put forward by \cite{Pad_2014} and \cite{Petro_2015}, if correct,
should materialise in an IceCube detection but this has not happened yet. At
present, in fact, IceCube has not identified any point sources and
therefore its signal remains unresolved, although the published upper limits are
still not ruling out the scenario described above \citep{Pad_2014}.

In this paper we aim to calculate the cumulative neutrino emission from {\it
  all} BL Lacs within the leptohadronic scenario for their $\gamma$-ray emission
  in order to see if BL Lacs {\it
  as a class} can indeed explain the IceCube detections. 
  By considering 
  only the neutrino emission produced in the blazar jet, 
 we calculate the neutrino background\footnote{We use ``background'' in the
  astronomical sense of total emission from a population of astrophysical
  objects. This includes resolved and unresolved sources, the contribution of
  the latter corresponding to the ``diffuse'' intensity generally referred to by
  high-energy physicists.}
  (NBG) from blazars. This is the most conservative estimate,
  since we do not take into account the `cosmogenic' neutrino production
 due to the propagation of escaping cosmic rays (CRs) in the interagalactic medium 
 \citep{Ahlers_2012}.
 As discussed below, calculating the NBG 
 from blazars requires a very detailed knowledge
of the blazar population in terms of \nup, $\gamma$-ray fluxes, redshift,
etc. All these parameters, and many more, are available in the simulations done
in a series of papers by Giommi, Padovani, and collaborators.

\cite{paper1} (hereafter Paper I) have proposed a new blazar paradigm where
blazars are classified into FSRQ and BL Lacs according to a varying mix of the
Doppler-boosted radiation from the jet, the emission from the accretion disc,
the broad-line region, and the light from the host galaxy. This is also based on
minimal assumptions on the physical properties of the non-thermal jet emission,
and unified schemes. These posit that BL Lacs and FSRQ are simply
low-excitation (LERGs)/Fanaroff-Riley (FR) I and high-excitation (HERGs)/FR II
radio galaxies with their jets forming a small angle with respect to the line of
sight \citep[e.g.][]{UP95}. They called this new approach the {\it blazar
  simplified view} (BSV). By means of detailed Monte Carlo simulations, Paper I
showed that the BSV scenario is consistent with the complex observational
properties of blazars as we know them from all the surveys carried out so far in
the radio and X-ray bands, solving at the same time a number of long-standing
issues.

In a subsequent paper \citep[][hereafter Paper II]{paper2} the Monte Carlo
simulations were extended to the \gr\ band (100 MeV -- 100 GeV) and found to
match very well the observational properties of blazars in the \fermi-LAT 2-yr
source catalogue \citep{fermi2fgl,fermi2lac} and the \fermi-LAT data of a sample
of radio selected blazars \citep{GiommiPlanck,RadioPlanck}.

\cite{paper3} (hereafter Paper III) considered the case of very high energy (VHE) 
emission ($E > 100$ GeV) extrapolating from the expectations for the GeV band,
and made detailed predictions for current and future Cherenkov facilities,
including the Cherenkov Telescope Array. Finally, \cite{Gio_Pad_2015} (hereafter
Paper IV) have extended the predictions below the sensitivity of current surveys
and estimated the contribution of blazars to the X-ray and \gr\ extragalactic
backgrounds. They found that the integrated light from blazars can explain 
the cosmic background at energies $\ga$ 10 GeV, and
contribute $\approx 40 - 70\%$ of the \gr\ diffuse radiation in the $0.1 - 10$
GeV band.

The purpose of this paper is to merge the simulation and theoretical
approaches by adding to the former a hadronic ``prior'' necessary for 
the neutrino emission, as detailed in Sect. \ref{ingredients}. 
As in papers I -- IV we use a $\Lambda$CDM cosmology with $H_0 = 70$ km
s$^{-1}$ Mpc$^{-1}$, $\Omega_m = 0.27$ and $\Omega_\Lambda = 0.73$
\citep{kom11}.

\section{Theoretical modelling}

Neutrino production is studied within a specific theoretical framework for
blazar emission where photohadronic interactions have an active role in shaping
the blazar SED, as detailed in \cite{Petro_2015}. In their model, the low-energy
emission of the blazar SED is attributed to synchrotron radiation of
relativistic electrons, whereas the observed high-energy (GeV -- TeV) emission
has a photohadronic origin. Under the assumption that proton acceleration to
high energies ($\sim 10^{16}-10^{17}$~eV) is also viable in blazar jets, the
production of charged pions is a natural outcome of photopion ($\pg$)
interactions between the relativistic protons and the internally produced
synchrotron photons. The decay of charged pions results in the injection of
secondary relativistic electron-positron pairs\footnote{We will commonly refer 
to them as electrons.}. It is the synchrotron radiation of the latter that
emerges in the GeV -- TeV regime, in contrast to the hadronic cascade scenario
for blazar emission \citep{mannheim91, mannheim93}, where the cascade emission
mainly contributes to the $\gamma$-ray regime\footnote{The cascade is initiated
  by photon-photon absorption of very high-energy $\gamma$-rays ($\gg$ TeV),
  which, in turn, are produced by synchrotron radiation of secondary pairs from
  pion decays and Bethe-Heitler ({\it pe}) pair production.}. As the
synchrotron self-Compton emission from primary electrons may also emerge in the
GeV -- TeV energy band, the observed $\gamma$-ray emission can be totally or
partially explained by photohadronic processes, depending on the specifics of
individual sources \citep{Petro_2015}. Since the luminosity of the $\pg$
component is directly connected to that of very high-energy ($\sim 2 - 20$~PeV)
neutrinos, our approach allows us to associate the observed $\gamma$-ray blazar
flux with the expected neutrino flux. The physical model we use has been
described, in more general terms, by \cite{dimi12} and \cite{dimi14}.

\section{Monte Carlo Simulations}\label{ingredients}

In the first two papers of this series we presented the principles on which the
BSV is built and mostly concentrated on the statistical properties of blazars,
such as distribution trends, average values of some important parameters, and
compared those with observed distributions and values in radio and X-ray
surveys. In \paperthree\ we tied our simulation to the absolute numbers of the
\fermi-2LAC catalogue and predicted the number of sources in the VHE band
taking into account the extragalactic background light (EBL) absorption. In
\paperfour\ we calculated the integrated flux from the entire population of
blazars of different types in the X-ray and \gr\ bands.

For the reader's convenience we briefly summarise here our Monte Carlo
simulations covering the radio through the
$\gamma$-ray bands, which are based on a number of ingredients, including: the
(radio) blazar luminosity function and evolution, a distribution of the Lorentz
factor of the electrons and of the Doppler factor, a synchrotron model, an
accretion disk component, the host galaxy, plus a series of $\gamma$-ray
constraints based on observed distributions estimated using simultaneous
multi-frequency data: the distribution of the ratio between high-energy and
low-energy hump fluxes, the dependence of the $\gamma$-ray spectral index on
$\nu_{\rm peak}^S$, and that of the $\gamma$-ray flux on radio flux
density. Sources are classified as BL Lacs, FSRQ, or radio galaxies based on
the optical spectrum, as in real surveys. Readers are referred to \paperone\ and
II for full details. We stress that no assumption has been made in the
simulations about the process responsible for the $\gamma$-ray emission.

As done in \paperthree\ the SEDs were extrapolated to the VHE band by using our
simulated \fermi\ fluxes and spectral indices and assuming a break at $E =
E_{\rm break}$ and a steepening of the photon spectrum by $\Delta \Gamma$. Our
adopted values were $E_{\rm break} = 100$ GeV and $\Delta \Gamma = 1$ and
$E_{\rm break} = 200$ GeV and $\Delta \Gamma = 0.5$, the latter being our
default choice. VHE spectra were attenuated using recent estimates of the EBL
absorption as a function of redshift \citep{Dom_2011}.

\subsection{The neutrino component}\label{sec:hadro}

The Monte Carlo simulations described above characterise fully the {\it photon}
emission from the BL Lac population. To get to the neutrinos, we need to add a
hadronic ``prior'', which is built on the knowledge gained by fitting the SEDs of six 
individual BL~Lac sources and their respective neutrino spectra in
\cite{Petro_2015}.

We model the observed differential neutrino plus anti-neutrino
($\nu+\bar{\nu}$) energy flux of all flavours ($F_{\nu}(E_{\nu})$) as

\eqb
F_{\nu}(\en) = F_0 \en^{-s} \exp\left(-\frac{\en}{E_0} \right),
\eqe

where the parameters to be defined are $E_0$
(characteristic energy), $s$ (spectral slope), and $F_0$ (normalisation). 

\subsubsection{Determination of $E_0$}\label{sec:hadro_eps}

 We set the characteristic energy $E_0$ equal to the peak energy of the
 neutrino spectrum $\ep$, which, in good approximation, can be written as

\eqb
\begin{multlined}
\ep(\delta, z, \ns) \simeq \frac{17.5 \ {\rm PeV}}{(1+z)^2}  \left(\frac{\delta}{10}\right)^2 \left(\frac{\ns}{10^{16}~{\rm Hz}}\right)^{-1},\\
E_0 = \ep~~~~~~~~~~~~~~~~~~~~~~~~~~~~~~~~~~~~~~~~~~~~~~~~~~~~~~~~~~\\
\end{multlined}
\label{eq:eps}
\eqe

where $\delta$ is the Doppler factor and $z$ is the source redshift
\citep[e.g.][]{dermer_07}. Here $\ns$ is the {\it observed} synchrotron
peak frequency. This relation is valid as long as protons energetic enough to
produce pions by interacting with the photons of the low-energy (synchrotron)
hump of the SED can be accelerated in the jet. We assume that this condition
is satisfied in all BL~Lacs {and discuss possible caveats in Sect.~\ref{sec:caveats}.

Equation \ref{eq:eps} requires as input $z$, $\delta$, and $\ns$. These are all
part of the output of the simulations so $E_0$ can be easily derived
source by source.

\subsubsection{Determination of $s$}\label{sec:hadro_s}

In \cite{Petro_2015} the neutrino spectra from six BL~Lac objects were
calculated self-consistently. It was shown that all neutrino spectra peaked at
$\sim \ep$ and $\en F_{\nu}(\en) \propto \en^{-s+1}$ with $\langle s \rangle \simeq
-0.35$, namely the neutrino spectra are relatively flat. This is not a matter of
fine tuning or coincidence as, in our scenario, the number density of the target
photons decreases with photon energy\footnote{This can be understood as follows:
  we assume that the proton distribution extends up to $\gamma_{\rm p, \max} \ga
  \gamma_{\rm p, th}$, where $\gamma_{\rm p, th}$ is the threshold energy for
  $\pg$ interactions with synchrotron photons of frequency $\ns$. Protons with
  $\gamma_{\rm p} < \gamma_{\rm p, th}$, which pion-produce on synchrotron
  photons with $\nu > \ns$, are responsible for the production of neutrinos with
  energies $\en < \ep$. Thus, the neutrino spectrum below its peak energy probes
  the soft part of the low-energy hump of the SED.}. Based on the range of
values found by \cite{Petro_2015}, the neutrino slope $s$ was drawn from a
Gaussian distribution centred at $-0.35$ with $\sigma = 0.12$.

\subsubsection{Determination of $F_0$}

The normalisation $F_0$ is given by

\eqb
F_0 = \frac{F_{\nu, \rm tot} \ep^{s-1}}{\int_{x_{\min}}^{\infty} dx \ x^{-s}e^{-x}},
\eqe

where $x\equiv E_{\nu}/\ep$ and $x_{\min}$ is the minimum normalised neutrino energy.
Because of the neutrino spectral shape, it is safe to set $x_{\min}=0$, and the
integral reduces to the $\Gamma$ function with argument $1-s$. In our approach,
the integrated neutrino flux is associated to the integrated $\gamma$-ray flux
as

\eqb
F_{\nu, \rm tot} = Y_{\nu \gamma} F_{\gamma}\left(> E_{\gamma} \right),
\eqe

where we chose $E_{\gamma}=10$~GeV\footnote{Our results are totally independent
  of this choice.}. By construction, $Y_{\nu \gamma}$ encloses all the
  information on the optical depth for photopion interactions and the proton
  luminosity. It has an upper limit, i.e. $Y_{\nu \gamma} \sim 3$, which is
obtained when synchrotron emission from $\pg$ pairs accounts for the whole
observed $\gamma$-ray flux \citep[e.g.][]{Petro_M_2015}. To reach this result,
we also made some simple assumptions about the pion production ratio and the
energy of the leptons produced from the pion decay chain.  However, there is no
lower limit on $Y_{\nu\gamma}$: $Y_{\nu \gamma} \ll 1$ simply implies a leptonic
origin for the $\gamma$-ray blazar emission.  
Application of the
  leptohadronic model to individual BL Lacs with different properties, such as
  SED and redshift, resulted in $Y_{\nu \gamma}$ values covering the range
  $0.1-2$ \citep{Petro_2015}. 

Based on the results of \cite{Petro_2015}, and assuming that $Y_{\nu\gamma}$ is
the same independently of BL Lac class (but see Sect. \ref{sec:LBL}) we adopt
two different options for the ratio $Y_{\nu \gamma}$:
\begin{enumerate}
 \item we assume a constant value for all BL~Lacs, $Y_{\nu \gamma}=0.8$, which
   is the average derived from the modelling of the six BL~Lacs;
\item we use a relation between $Y_{\nu \gamma}$ and the observed $\gamma$-ray
  luminosity above $10$~GeV in erg s$^{-1}$, $L_{\gamma}(> 10$ GeV), i.e.  $\log Y_{\nu
    \gamma} = -0.58 \times \log L_{\gamma}(> 10$ GeV$) + 26.3$, with $Y_{\nu \gamma} \le
  3$. This is a simple fit to the values in \cite{Petro_2015} (see their
  Fig. 11) motivated by the possible trend found in the sample of six
  BL~Lacs. In practice, this translates to $Y_{\nu \gamma} = 3$ for 
  $L_{\gamma}(> 10$ GeV) $< 3 \times 10^{44}$ erg s$^{-1}$, that is for basically all LBL and
  $\sim 80\%$ of HBL.
  \end{enumerate}

Combining all of the above, we derive the final expression for the observed
neutrino flux expected from each BL~Lac object ($E^2 dN/dE$ units):

\eqb
\begin{multlined}
\en F_{\nu}(\en) = \\ \frac{Y_{\nu \gamma} F_{\gamma}(> {\rm 10~GeV})}{\int_{x_{\min}}^{\infty} dx \ x^{-s} e^{-x}}\left(\frac{\en}{\ep}\right)^{-s+1}\exp\left(-\frac{\en}{\ep}\right).
\label{eq:flux}
\end{multlined}
\eqe

The neutrino fluxes at various energies were then calculated based on this
spectrum, which in turn depends on $\ep$, $s$, and $Y_{\nu \gamma}$,
and the $\gamma$-ray flux and power from the simulations. Since $\ep$
is fully determined by eq.~\ref{eq:eps} and $s$ covers a narrow range, we are
left only with $Y_{\nu \gamma}$ as a possible ``tuneable'' parameter (see
Sect. \ref{sec:predictions}).

Finally, the total NBG was computed as $B =\int_{F_{min}}^{F_{max}}
F~\frac{dN}{dF}~dF$ where $\frac{dN}{dF}$ are the differential number counts and
$F_{min}$ and $F_{max}$ are the fluxes over which these extend. Since the
IceCube data are ``per neutrino flavour'' our numbers, which refer to all
neutrino flavours, were divided by three\footnote{The neutrino flux produced at
  the source contains neutrinos of different flavours with an approximate ratio
  $F_{\nu_{\rm e}}:F_{\nu_{\mu}} : F_{\nu_{\tau}} = 2 : 1 : 0$. However, by the
  time they reach Earth their ratio will have changed to $F_{\nu_{\rm
      e}}:F_{\nu_{\mu}} : F_{\nu_{\tau}} = 1 : 1 : 1$ due to neutrino
  oscillations \citep{learned95}, 
 as indeed observed \citep{Aartsen_2015}.
  }. 
  Ten simulations were run and the average was
then calculated to smooth out the ``noise'' inherent to the Monte Carlo process.

\begin{figure}
\includegraphics[height=8.6cm]{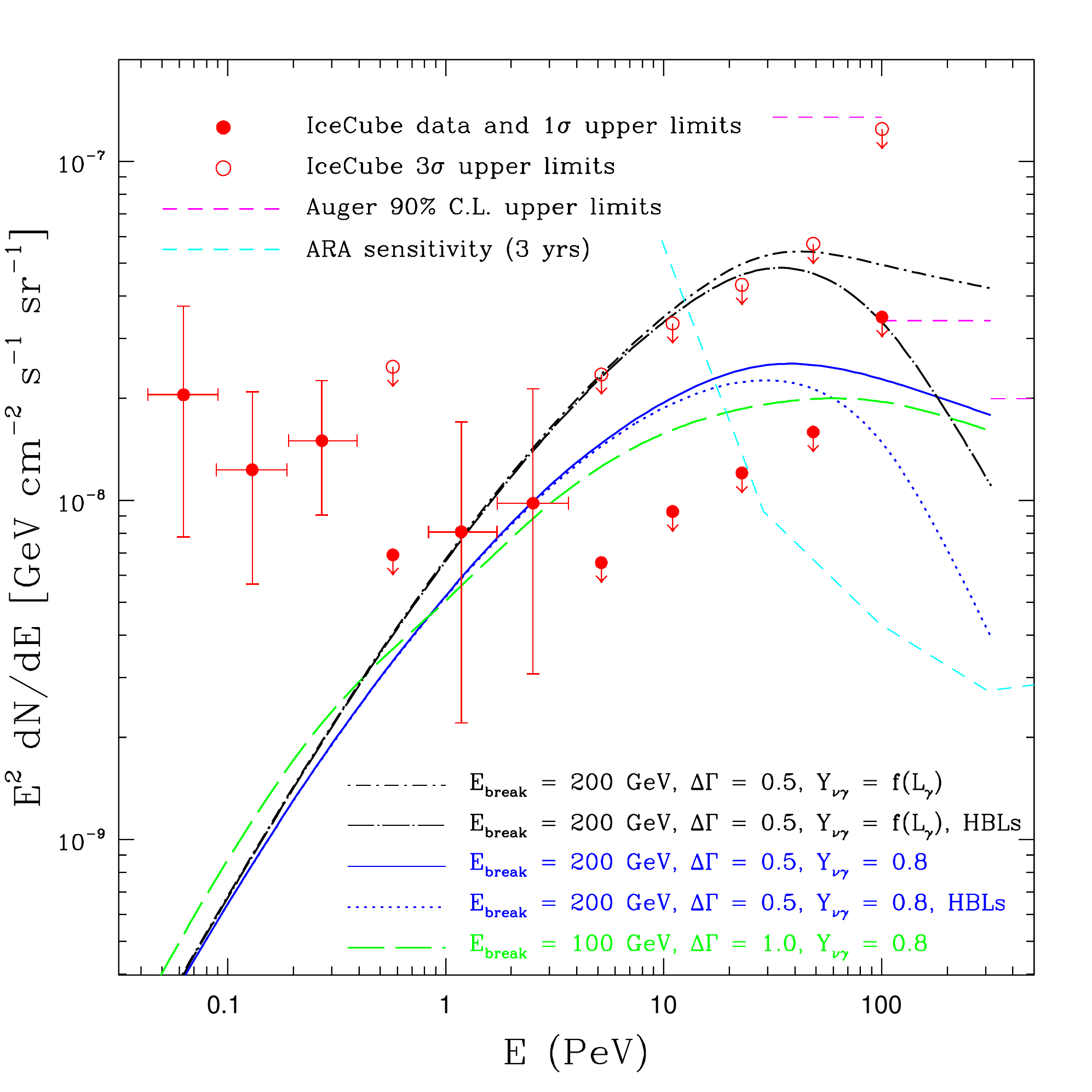}
\caption{The predicted neutrino ($\nu+\bar{\nu}$) background per neutrino
  flavour from BL Lacs. Different lines correspond to different assumptions
  (starting from the top): $Y_{\nu\gamma}$ anti-correlated with $L_{\gamma}(> 10$ GeV)
  for all BL Lacs (black dot short-dashed line) and HBL (black dot long-dashed
  line); $Y_{\nu\gamma} = 0.8$ and $E_{\rm break} = 200$ GeV, $\Delta \Gamma =
  0.5$, for all BL Lacs (blue solid line) and HBL (blue dotted line);
  $Y_{\nu\gamma} = 0.8$ and $E_{\rm break} = 100$ GeV, $\Delta \Gamma = 1$, for
  all BL Lacs (green long dashed line). All lines correspond to the mean value
  of ten different simulations. The (red) filled points are the data points from
  \protect\cite{ICECube14}, while the open points are the $3\sigma$ upper
  limits. The upper (magenta) short dashed lines represents the 90\% C.L. upper
  limits from \protect\cite{Auger2015} while the lower (cyan) short dashed line
  is the expected three year sensitivity curve for the Askaryan Radio Array
  \protect\citep{ARA_2012}.}
\label{fig:background}
\end{figure}

\section{The neutrino background from BL Lacs}\label{sec:NBG}

Fig.~\ref{fig:background} shows our results, with different lines
corresponding to different assumptions. Namely, and starting from the top:
$Y_{\nu\gamma}$ anti-correlated with $L_{\gamma}(> 10$ GeV) for all BL Lacs (black dot
short-dashed line) and HBL (black dot long-dashed line); $Y_{\nu\gamma} =
0.8$ and $E_{\rm break} = 200$ GeV, $\Delta \Gamma = 0.5$, for all BL Lacs
(blue solid line) and HBL (blue dotted line); $Y_{\nu\gamma} = 0.8$ and
$E_{\rm break} = 100$ GeV, $\Delta \Gamma = 1$, for all BL Lacs (green long
dashed line). The (red) filled circles are the data points from
\cite{ICECube14}, while the open points are the $3\sigma$ upper limits,
derived from the $1\sigma$ ones by simply scaling them by a factor $\sim 3.6$
\citep{geh86}. The upper (magenta) short dashed lines represents the 90\%
C.L. upper limits from \cite{Auger2015} while the lower (cyan) short dashed
line is the expected three year sensitivity curve for the Askaryan Radio
Array \citep{ARA_2012}.

A few points can be made about this figure:

\begin{enumerate}

\item BL Lacs as a class can easily explain the whole NBG at high-energies ($\ga
  0.5$ PeV) while they do not contribute much ($\sim 10\%$) at low-energies
  ($\la 0.5$ PeV);

\item HBL make up the bulk of the NBG in all cases up to $\approx 30$ PeV, where
  they start to contribute less and less. This is where LBL take over, due to
  their lower $\nu_{\rm peak}^S$ and therefore larger values of neutrino peak
  energy, given the inverse dependence between $\nu_{\rm peak}^S$ and $\ep$
  (eq. \ref{eq:eps}). Note that HBL manage to dominate the neutrino output
  despite their small fraction ($\sim 5\%$) because of their relatively high
  $\gamma$-ray, and therefore neutrino, fluxes and powers;
  
\item there is very little difference between the $E_{\rm break} = 200$ GeV,
  $\Delta \Gamma = 0.5$ and the $E_{\rm break} = 100$ GeV, $\Delta \Gamma = 1$
  cases. This is due to the fact that we relate the neutrino flux to
  $F_{\gamma}(> {\rm 10~GeV})$ and most of the $\gamma$-ray flux is at lower
  energies. Thus, the neutrino flux in the latter case is only slightly smaller
  than in the former;
  
\item The difference between the constant and
  varying $Y_{\nu\gamma}$ case is small up to $\en \sim 1 - 2$ PeV. 
  However, at $\en \ga 5$ PeV only the constant $Y_{\nu\gamma}$ case is
  fully consistent with the IceCube data, while the varying $Y_{\nu\gamma}$
  scenario is barely within the $3\sigma$ upper limits. Moreover, current Auger 
  upper limits already rule out our varying
  $Y_{\nu\gamma}$ scenario, while future ones will better probe the high-energy
  tail. The Askaryan Radio Array, by sampling the $\ga 10$ PeV range with high
  sensitivity, will further constrain our model, together with the Antarctic
  Ross Ice Shelf Antenna Neutrino Array \citep[ARIANNA,][]{ARIANNA_2015}, which
  will cover the 100 PeV -- 10 EeV energy range;

\item if our model is correct, IceCube should see more $\en > 1$ PeV, and even $\en
  > 2$ PeV, events; alternatively, a non-detection will put strong constraints
  on $Y_{\nu\gamma}$ (see Sect. \ref{sec:predictions}).

\end{enumerate}

\begin{figure}
\includegraphics[height=8.6cm]{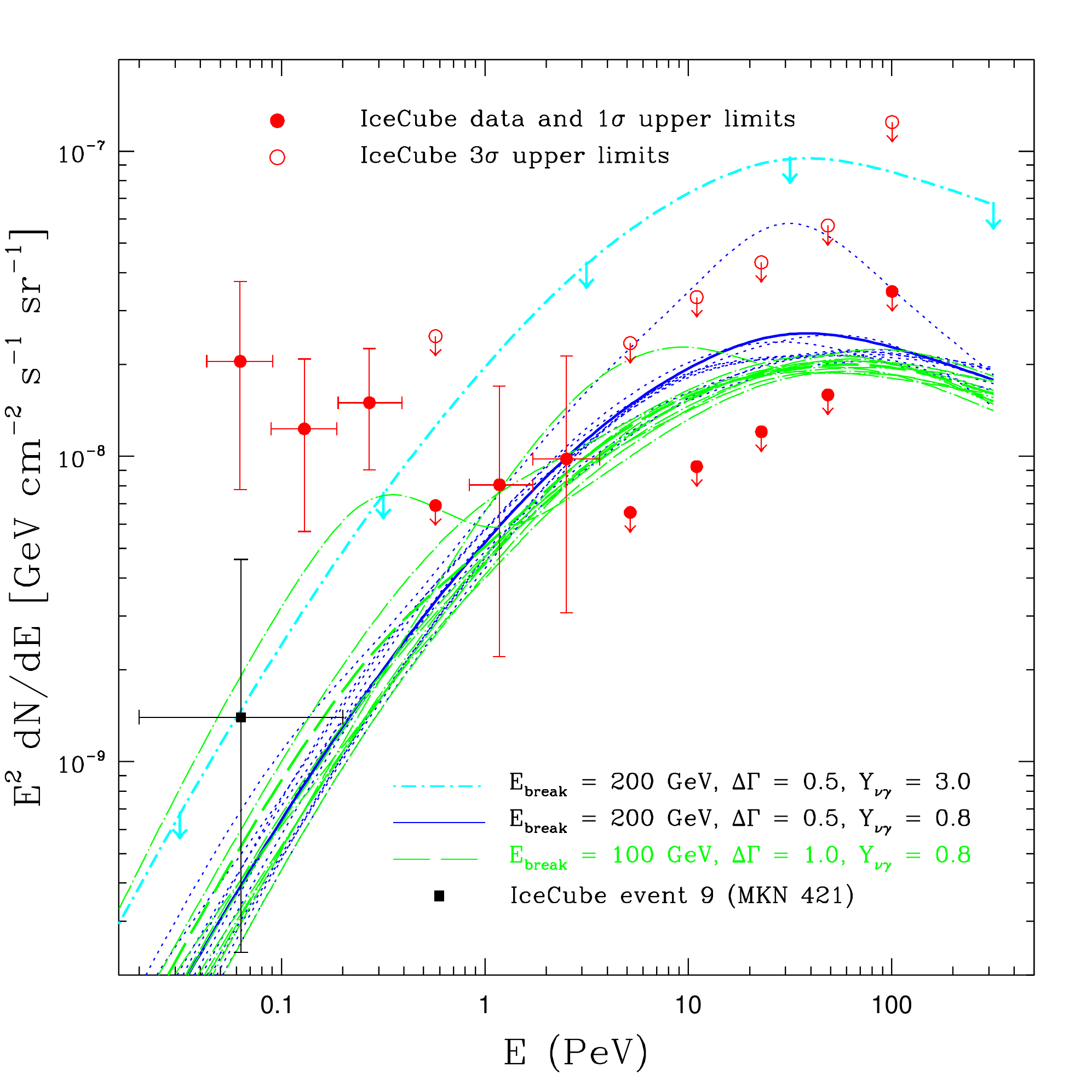}
\caption{The predicted neutrino ($\nu+\bar{\nu}$) background per neutrino
  flavour from BL Lacs for three different cases and showing the range of
  results from ten different simulations. Different lines correspond to
  different assumptions (starting from the top): $Y_{\nu\gamma} = 3$ (the
  maximum theoretical value: cyan dot short-dashed line) and $E_{\rm break} =
  200$ GeV, $\Delta \Gamma = 0.5$; $Y_{\nu\gamma} = 0.8$ and $E_{\rm break} =
  200$ GeV, $\Delta \Gamma = 0.5$: average value (solid blue line) and
  individual runs (dotted blue lines); $Y_{\nu\gamma} = 0.8$ and $E_{\rm break}
  = 100$ GeV, $\Delta \Gamma = 1$: average value (long dashed green line) and
  individual runs (dot long-dashed green lines). The (red) filled points are the
  data points from \protect\cite{ICECube14}, while the open points are the
  $3\sigma$ upper limits. The black square is the neutrino flux of IceCube event
  9, from \protect\cite{Pad_2014}, who have tentatively associated it with MKN
  421, converted to ``background'' units (i.e. divided by $4\pi$).}
\label{fig:background_2}
\end{figure}

Given the above discussion, from now on we refer to the $Y_{\nu\gamma} = 0.8$
and $E_{\rm break} = 200$ GeV, $\Delta \Gamma = 0.5$ case as our ``benchmark''
case.

\cite{Gio_Pad_2015} have shown that if the Monte Carlo simulations are modified
to include the strong dependence of \nup\ on radio power postulated by the
blazar sequence \citep{fossati98} the agreement with the observational data
disappears, as the predicted $\gamma$-ray background above a few GeV turns out
to be far in excess of the observed value. A similar thing happens for the
predicted NBG, which, for example, turns out to be more than two orders of
magnitude above the IceCube data at $\en \sim 1$ PeV.

Fig. \ref{fig:background} gives a feeling for the average NBG but not for the
dispersion intrinsic to our simulation process. This is shown in
Fig. \ref{fig:background_2}, where the results from the ten different
simulations are displayed individually. Different lines correspond to different
assumptions. Namely, our benchmark case: average value (solid blue line) and
individual runs (dotted blue lines); $Y_{\nu\gamma} = 0.8$ and $E_{\rm break} =
100$ GeV, $\Delta \Gamma = 1$: average value (long dashed green line) and
individual runs (dot long-dashed green lines). The top line (cyan dot
short-dashed line) represents the case with $E_{\rm break} = 200$ GeV, $\Delta
\Gamma = 0.5$ but $Y_{\nu\gamma} = 3$. As this is the maximum value expected in
our model for this
parameter \citep[e.g.][]{Petro_M_2015} this curve represents the largest NBG
from BL Lacs. The black square is the neutrino flux of IceCube event 9
\citep{Pad_2014} converted to ``background'' units (i.e. divided by $4\pi$).

\begin{figure}
\includegraphics[height=8.5cm]{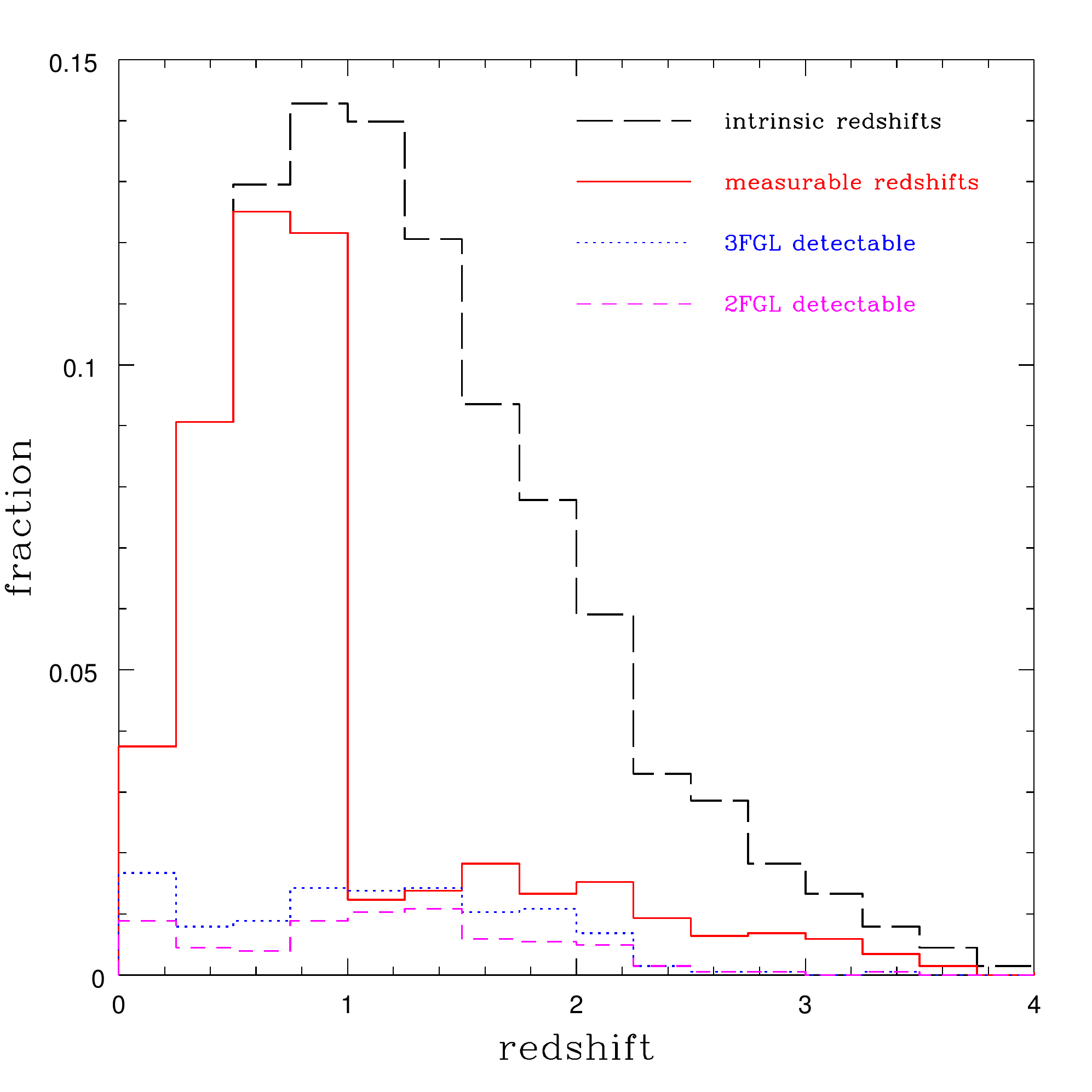}
\caption{The redshift distribution for the BL Lacs contributing $\sim 95\%$ of
  the background associated with the benchmark case at 1 PeV (black long-dashed
  line) and for those with a measurable redshift (red solid line). The sources
  detectable by the 3FGL (blue dotted line) and 2FGL (magenta short-dashed line)
  catalogues are also indicated.}
\label{fig:nz}
\end{figure}

A few points can be made about this figure:

\begin{enumerate}

\item the dispersion amongst the ten different simulations for the two cases is
  typically not very large ($\la 0.3$ dex), with only $1 - 2$ of them being
  clear outliers. In all cases this is due to a {\it single} source, which
  has a neutrino flux $\sim 20 - 45$ times higher than the source with the second largest
  flux, at the energy where a ``bump'' can be seen in the
  curves. This has to be expected given the nature of the simulations but these
  outliers happen only $\sim 10 - 20\%$ of the time;

\item the presence of these outliers can explain at least some of the low-energy
  associations made by \cite{Pad_2014}. The neutrino flux of IceCube event 9,
  which has been tentatively associated by these authors with MKN 421, is in
  fact fully compatible with one of the runs. Moreover, the number of events
  predicted by some of these outliers at low energies ($\en < 2 $ PeV) is about
  twice as large as the mean of the ten simulations;
  
\item the curve for the case with $Y_{\nu\gamma} = 3$ defines the region of
  parameter space allowed by our model. At high energies
  this is well above the IceCube $3\sigma$ upper limits, which means that this
  upper bound is clearly ruled out by current data, and therefore that a
  leptonic component must be present in the $\gamma$-ray emission of blazars.

\end{enumerate}

\section{Discussion}\label{sec:discuss}

\subsection{Characterising the NBG sources}

As is normally the case for astrophysical backgrounds, most of the NBG
is produced by a small fraction of the population. In our case, $\sim 0.5\%$ of
the BL Lacs make $\sim 95\%$ of the background associated with the benchmark
case at 1 PeV, which is where our predictions are quite close to the IceCube
data. We now focus on this sub-sample and describe some of its properties.

Fig. \ref{fig:nz} shows the redshift distributions for the whole
sub-sample\footnote{We note that this is not significantly different from the
  redshift distribution of the entire sample.} (dashed line) and for the subset
of BL Lacs with measurable redshift (red solid line). Note that the latter make 
up $\sim 50\%$ of the sub-sample, with the other $50\%$ consisting of BL Lacs 
with {\it no} measurable redshift. These are BL Lacs whose
maximum equivalent width (EW) is $\la 2$ \AA, or for which the non-thermal light
is at least a factor 10 larger than that of the host galaxy, and are therefore
deemed to have a redshift which cannot be typically measured (\paperone). As
expected, the two distributions are quite different with $\langle z \rangle \sim
1.3$ and $\sim 0.9$ respectively. This is due to the fact that the sources with
no measurable redshift are mostly those with the optical spectra swamped by
non-thermal emission, which tend to have a more powerful jet and therefore are
on average at larger intrinsic redshift. 

The fraction of sources in the sub-sample detectable by the 3FGL {\it Fermi}
catalogue \citep{3FGL} is $\sim 11\%$. However, these make $\sim 50\%$ of the
overall background associated with the benchmark case, i.e. $\sim 15\%$ of the
NBG. We also find that $\sim 7\%$ of the sources would be detectable by the 2FGL
\citep{fermi2fgl}, making $\sim 40\%$ of the background associated with the
benchmark case, i.e. $\sim 12\%$ of the NBG. These numbers are consistent with
the preliminary results of \cite{gluse_2015}, who find no evidence of neutrino
emission and a maximal contribution from {\it Fermi} 2LAC \citep{fermi2lac}
blazars $\sim 20\%$\footnote{This is however derived assuming an $E^{-2.46}$
  spectrum, which is much softer than our predictions.}.
  
  The (intrinsic) redshift distributions of the sources detectable by the 3FGL
  (blue dotted line) and 2FGL (magenta short-dashed line) catalogues are also 
  shown in Fig. \ref{fig:nz}.

\subsection{Possible caveats}\label{sec:caveats}

\subsubsection{LBL}\label{sec:LBL}

The inclusion of the hadronic ``prior'' into the Monte Carlo simulations is
based upon the results derived from individual SED fitting of HBL (Petropoulou
et al. 2015). The application of our model to LBL, therefore, could not be
straightforward. Thus, we discuss here possible caveats.

A direct application of eq.~\ref{eq:eps} to LBL requires that: (1) the photons
of the low energy SED component are the main targets for photopion interactions;
(2) acceleration of protons to ultra-high energies (UHE), e.g. $10^{19} -
10^{20}$ eV, takes place in the jets of these sources. These assumptions, in
turn, suggest that specific conditions should prevail in the emitting region of
individual LBL, in order to explain their $\gamma$-ray emission in terms of
photohadronic processes. In particular, if the $\gamma$-ray emission is
synchrotron radiation of UHE secondary pairs (similarly to HBL), then the
emitting region should contain very weak magnetic fields, e.g. $B \ll 3\mu$G
\citep[see also eq.~9 in][]{Petro_M_2015}. Alternatively, the $\gamma$-ray
emission of LBL  could be the result of a hadronic cascade
\citep{mannheim91,mannheim92}. If the above conditions can be realised in the
jets of LBL, then application of eq. \ref{eq:eps} shows that the LBL
contribution to the NBG takes place at high energies, i.e. $> 100$~PeV (see
Fig.~\ref{fig:background}).

One could relax, however, assumption (2), if the targets for the photopion
interactions were synchrotron photons with frequencies above $\nu_{\rm
  peak}^{S}=10^{13}$~Hz \citep[see generic SED in Fig. 8 in][]{Petro_M_2015}.
In this case, application of eq.~\ref{eq:eps} to LBL would also result in the
production of neutrinos with energies similar to those expected from
HBL\footnote{A preliminary application to Ap~Librae results in $Y_{\nu
    \gamma} \sim 0.5$ and a neutrino peak energy of several PeV.}. In this
scenario, the neutrino emission from both HBL and LBL sources is expected to
have a similar peak energy.

To test the robustness of the results shown in Figs. \ref{fig:background} and
\ref{fig:background_2} we therefore artificially shifted the $\nu_{\rm
  peak}^{S}$ distribution of LBL so that it had the same mean as HBL, using
the same value of $Y_{\nu \gamma}$. We then repeated the Monte Carlo
simulations and found that the total NBG flux increased only by $\simeq 15\%$
overall. This is due to the fact that LBL have much smaller $\gamma$-ray, and
therefore neutrino, fluxes than HBL. It then follows that our results on the
NBG do not depend on a detailed modelling of LBL and are therefore robust.

\subsubsection{Energetics}\label{sec:energetics}

Leptohadronic models are known to predict relatively high jet powers, driven by
their relativistic proton content \citep[e.g.][and references
  therein]{zdz_15,Petro_2015}. Indeed, the estimated jet power for the six HBL
to which \cite{Petro_2015} applied their leptohadronic model, and which lie at
the basis of our calculations, is quite high ($L_{\rm jet} \sim 2 \times 10^{48}
- 8 \times 10^{49}$ erg s$^{-1}$ [see their Tab. 4]).

One could compare $L_{\rm jet}$ with the Eddington luminosity, $L_{\rm Edd}$, of
those sources. Assuming a typical black hole mass of $3 \times 10^8~M_{\odot}$
\citep{Plotkin_2011}, this would be $\sim 3.8 \times 10^{46}$ erg s$^{-1}$,
which, in turn, would imply that the jet power is $\ga 50$ (and up to $\sim
2,000$) times larger. These large $L_{\rm jet}/L_{\rm Edd}$ ratios are
not necessarily alarming, for the following reasons: 1. out of the six BL Lacs
modelled in \cite{Petro_2015} only for MKN 421 is there an estimate for the
black hole mass \citep{Sbarrato_2012} (which happens to be equal to the value we
have assumed), which means that a comparison against an uncertain $L_{\rm Edd}$
is not very informative; 2. \cite{Ghis_2014}, by applying a {\it leptonic} model
to a large sample of mostly FSRQ, have derived at low powers $L_{\rm jet}
\approx L_{\rm Edd}$, with some sources reaching $L_{\rm jet}/L_{\rm Edd}
\approx 30$. At least some of the sources studied by \cite{Petro_2015}
(including MKN 421) lie close to the upper range of that distribution; 3. the
total jet power does not need to be limited by $L_{\rm Edd}$, as is observed,
for example, in the case of gamma-ray bursts, which are super-Eddington by huge
amounts \citep[$\approx 10^{10}$; e.g.][]{Piran_2004}.

A more meaningful comparison would be that of $L_{\rm jet}$ with $\dot M c^2$,
where $\dot M$ is the accretion rate. One in fact expects $L_{\rm jet} \sim
\epsilon_j \dot M c^2$ with $ \epsilon_j \la 1.5$ \citep[][and references
  therein]{zdz_15}. In the case of radiatively efficient accretors, the
accretion rate is normally derived from the relation $\dot M c^2 = L_{\rm
  disk}/\eta$, where $\eta$ is the radiative efficiency. $L_{\rm disk}$, in
turn, is estimated through the broad emission lines, which are used, via a
template, to derive the luminosity of the entire broad line region ($L_{\rm
  BLR}$). The latter is therefore a proxy for the accretion disk luminosity,
modulo a covering factor, which is usually taken to be $\sim 0.1$ \citep[which
  means $L_{\rm disk} \sim 10 L_{\rm BLR}$:][and references
  therein]{Ghis_2014}. None of the six sources studied by \cite{Petro_2015}
shows broad lines, while only two sources display narrow lines in their
spectra. In fact, our sources are most likely LERGs and therefore radiatively
inefficient accretors. In short, it is not possible to derive the accretion
power of these sources, since they do not exhibit broad line features. Most
importantly though, we could not relate $L_{\rm disk}$ to $\dot M c^2$ because
the simple relation, which is valid for radiatively efficient accretors, breaks
down for LERGs \citep[][and references therein]{Sbarrato_2014}. The point
remains that we do not expect high values of $\dot M$, as otherwise these
sources could not be LERGs.

If our model will be confirmed by future IceCube detections, this
will simply mean that the current picture of accretion in broad-lined,
efficient accretor AGN cannot be applied to HBL.  

As a final point, we note that the energetics for BL Lacs in the proton
synchrotron scenario are more reasonable than in the leptohadronic one
\citep{dimi14} but in that case the neutrino flux from BL Lacs would peak at
$\en \sim 200$ PeV and would be much lower than in our model.

It could be argued that, given the high powers in relativistic protons, proton-proton 
($pp$) collisions with the background (thermal) protons in the inner jets of blazars 
could result in a substantial production of neutrinos. While a detailed study of this 
process goes beyond the scope of this paper, we note that, having a neutrino flux 
from the $pp$ mechanism equal to that from the $\pg$ interaction at $\sim 1$ PeV, 
would require thermal proton number densities $\approx 10^8$ cm$^{-3}$. These, in turn, 
would translate into a (thermal) proton energy density $\sim 1 - 2$ orders of 
magnitude larger than that derived by \cite{Petro_2015} and a
correspondingly larger $L_{\rm jet}$. We thus contend that this scenario is not plausible 
\citep[see also][]{atoyandermer03}. 

\subsubsection{FSRQ}\label{sec:fsrq}

The model we used for including neutrino emission in the Monte Carlo
simulation has not been applied to FSRQ, which we therefore excluded from our
calculations. A proper treatment of the neutrino emission from individual FSRQ
is required in order to make robust predictions about their cumulative
contribution, but this lies outside the scope of the present study. 
Recently, \cite{dermer_2014} and \cite{muraseinouedermer14} have calculated the 
neutrino background from FSRQ based on the blazar sequence and assuming a leptonic 
origin of the blazar $\gamma$-ray emission. They have taken into account photopion 
interactions with the non-thermal emission from the jet and the external photon fields 
(BLR and dusty torus). The latter have been shown to provide most of the contribution 
to the total neutrino output.

Here instead, we calculated the NBG from FSRQ assuming that they fall within the
same scenario of BL Lacs, an assumption that has no astrophysical basis. We find
that their contribution to the NBG is somewhat energy dependent but overall only
$\approx 30\%$ that of BL Lacs. We caution the reader that this estimate
neglects the role of external photon fields in the photopion interactions and,
therefore, in the final neutrino output. As such, this contribution should
  be considered as a lower limit.

\subsubsection{Masquerading BL Lacs}\label{sec:masq}

\paperone\ and II discussed the existence of sources, which appeared ``BL
Lac-like'' only because their emission lines were heavily diluted by strong
non-thermal emission. These objects, which are therefore ``masquerading" BL
Lacs and are in fact misclassified FSRQ, make up a very small fraction
($\sim 2\%$) of our BL Lacs. However, as is the case for HBL, because of
their relatively high $\gamma$-ray, and therefore, neutrino fluxes, their
contribution to the NBG is $\approx 60\%$ that of all BL Lacs. Our
calculations on the neutrino emission were made under the assumption that
only photons produced internally in the jet, i.e. synchrotron photons, are
the targets for photopion interactions. In masquerading BL~Lacs any line- or
blackbody-like emission external to the jet is hidden below the non-thermal
synchrotron emission that is produced within the jet. We argue that our model
is also applicable to these sources, as long as $u'_{\rm syn} \ga u'_{\rm
  ex}$, where $u'_{\rm syn}$ and $u'_{\rm ex} \simeq \Gamma^2 u_{\rm ex}$ (where
  $\Gamma$ is the Lorentz factor) are
the energy densities of synchrotron and external photons, respectively, as
measured in the rest-frame of the emitting region. Obviously, our argument
becomes even stronger if the boosting of the external energy density becomes
less efficient (e.g. the non-thermal emitting region is located much further out than 
the external photon field region).

\subsubsection{Flaring sources}

One of the defining characteristics of blazars is their large variability known
to occur in all parts of the electromagnetic spectrum. The specific amount
depends on the energy band and on the blazar type, with values ranging from a
factor of a few in the radio band up to a factor 10,000 at GeV energies
\citep[e.g.][]{aharonian07,giommi15}. The amount of variability of the neutrino
flux in blazars and how this correlates with $\gamma$-ray variability is not
known. As the number of astrophysical neutrino events detected so far is still
very small, one could speculate that large flaring events from one or a few of
the brighter sources in Fig. \ref{fig:background_2} may produce a fair fraction
of the observed events. This is not straightforward though, and requires
detailed modelling of flaring sources \citep[e.g.][]{reimer05}. A more recent
application to the flaring MKN~421 also suggests that an accumulation of flaring
events from an individual source may be needed in order to have a
statistically significant neutrino detection (Petropoulou et al., in
preparation). Here we consider the emission arising from the entire
population of BL Lacs and assume that flares from single sources will be
diluted by the integrated emission from all blazars.

\subsection{Comparison with previous results}

The idea of blazars being sources of high-energy neutrinos dates back to long
before the detection of sub-PeV neutrinos with IceCube \citep[]{ICECube13} and
has since been explored in a number of studies \citep[e.g.][]{mannheim95,
  halzen97, mueckeetal03,tav15,kis14,dermer_2014,muraseinouedermer14}. 
  Our approach has some similarities but
many differences with previous work, as detailed below.

\subsubsection{Similarities}

  \begin{enumerate} 
   \item the BL Lac $\gamma$-ray emission has a (photo)hadronic origin (at least for
   the models presented in Fig. \ref{fig:previous}); 
   \item in BL Lacs the targets for photopion interactions are the low-energy
     synchrotron photons.
  \end{enumerate} 

\subsubsection{Differences}

 \begin{enumerate}
  \item we use as a starting point the knowledge gained from detailed SED
    fitting of BL Lacs instead of using a generic neutrino spectrum (e.g. 
    \citealt{mannheimetal01, kis14, muraseinouedermer14}). By establishing
    a connection between the $\gamma$-ray and neutrino emission for each source
    (see eqs.~\ref{eq:eps} -- \ref{eq:flux}), we are able to assign to {\it each}
    simulated BL Lac in the Monte Carlo code a unique neutrino spectrum.  We
    then calculate the NBG by summing up the fluxes of all sources in each
    energy bin;
 \item for the calculation of the NBG we do not normalise {\sl a priori} a
   generic neutrino spectrum to the extragalactic $\gamma$-ray background (EGB)
   \citep[e.g.][]{mannheim95, mueckeetal03}. In fact, we do not need to, as our
   simulation naturally {\sl reproduces} the observed EGB above 10 GeV
   \citep{Gio_Pad_2015};
  \item the NBG spectrum is not {\sl a priori} normalised to the IceCube
    observations \citep[e.g.][]{tav15}. Instead, for a specific choice of
    $Y_{\nu \gamma}$, which is the only tuneable parameter in our framework, we
    compare our model predictions with the IceCube data;
  \item the maximum proton energy is taken to be a few times larger than the
    threshold energy for photopion interactions with the peak energy synchrotron
    photons of the low-energy hump. This is usually lower than the values used
    in previous studies \citep[e.g.][]{halzen97, mueckeetal03}, which also
    explains the difference in the peak energies of the NBG;
  \item the $\gamma$-ray emission of individual BL Lacs in our approach is a
    combination of synchrotron radiation emitted by electron-positron pairs
    produced through $\pi^{\pm}$ decay and synchrotron self-Compton from primary
    electrons. The cascade emission initiated by $\pi^0$ $\gamma$-rays has a
    negligible effect in the formation of the blazar SED. This is in contrast to
    previous studies, where the blazar $\gamma$-ray emission is explained either
    as proton synchrotron radiation (e.g. \citealt{mueckeetal03}) or as
    cascade emission (e.g. \citealt{halzen97, kis14}).
 \end{enumerate}
 
\begin{figure}
\includegraphics[height=8.5cm]{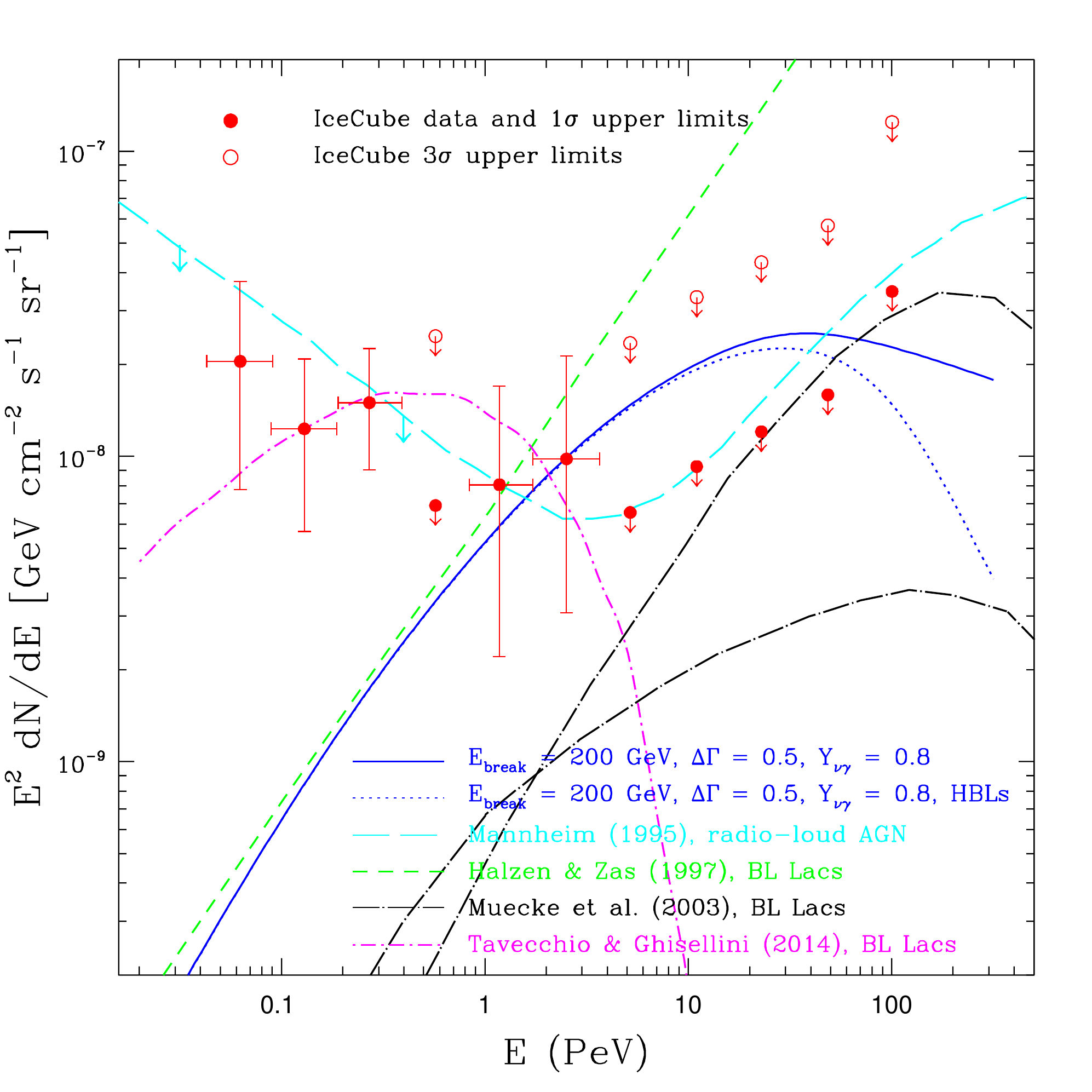}
\caption{The predicted neutrino background per neutrino flavour for
  $Y_{\nu\gamma} = 0.8$ and $E_{\rm break} = 200$ GeV, $\Delta \Gamma = 0.5$,
  for all BL Lacs (blue solid line) and HBL (blue dotted line) compared to
  previous results. Namely, in chronological order: \protect\cite{mannheim95}
  (long dashed cyan line; upper limits at low energies), \protect\cite{halzen97}
  (short dashed green line), \protect\cite{mueckeetal03} (dot long-dashed black
  lines: LBL, upper curve; HBL, lower curve), and \protect\cite{tav15} (dot
  short-dashed magenta line). The (red) filled points are the data points from
  \protect\cite{ICECube14}, while the open points are the $3\sigma$ upper
  limits. See text for details.}
\label{fig:previous}
\end{figure}

\subsubsection{Detailed comparison}  
  
Fig.~\ref{fig:previous} compares the predicted neutrino background for our
benchmark case for all BL Lacs (blue solid line) and HBL (blue dotted line) with
some of the previous results. In chronological order, these are:
\cite{mannheim95} (long dashed cyan line, upper limits at low energies),
\cite{halzen97} (short dashed green line), \cite{mueckeetal03} (dot long-dashed
black lines), and \cite{tav15} (dot short-dashed magenta line). The two curves
from \cite{mueckeetal03} represent the {\it maximum} contribution expected from
LBL (upper curve) and HBL (lower curve), respectively.
A few things about Fig.~\ref{fig:previous} are worth
mentioning:
\begin{enumerate}
 \item the model by \cite{mannheim95} at first glance is the one that best
   describes the IceCube data. This, taken at face value, would imply that
   radio-loud AGN explain the entire NBG, something that contradicts the
   preliminary IceCube results of \cite{gluse_2015}, who find a maximal
   contribution from {\it Fermi} 2LAC \citep{fermi2lac} blazars $\sim
   20\%$. However, since it gives only upper limits at low energies, it could be
   still reconciled with the data.  This model has a very different shape as
   compared to the others because it includes {\it two} hadronic components,
   i.e. a low-energy soft one ($\en \la 2$~PeV), produced through $pp$
   collisions of the escaping CRs from the blazar jet with the
   ambient medium, and a high-energy flat one ($\en \ga 2$~PeV), related to
   $\pg$ interactions of CRs with the synchrotron photons in the blazar jet;
\item the model by \cite{halzen97}, although very close to ours at low energies,
  lies above the 3$\sigma$ upper limits at $\en \ga 5$ PeV, while the sum of the
  two curves by \cite{mueckeetal03} remains consistently below the IceCube data.
  Although the model curve of \cite{tav15}
  passes through the data points, this is by construction,
  i.e. the NBG was {\sl a priori} normalised to the IceCube data. Moreover, this model
 might also contradict the IceCube results of \cite{gluse_2015} mentioned 
 above (pending the different 
 spectral shapes);
\item apart from the \cite{tav15} model and the low-energy ($\en < 2$~PeV) part
  of the \cite{mannheim95} model, the NBG spectrum below its peak is relatively
  hard in all other models shown in Fig.~\ref{fig:previous}. This is related to
  the assumed proton power-law index and the spectrum of target photons, which
  are both similar among the models. The diversity of the peak energy of the NBG
  spectrum, on the other hand, reflects mainly the different model assumptions
  on the maximum proton energy in blazars.
\end{enumerate}
    
\subsection{Cosmic rays}

Neutrons that are produced in photopion interactions are an effective means of
CR escape from the system. They are unaffected by its magnetic field, their
decay time is high enough to allow them to escape freely before reverting into
protons \citep{kirkmast89,begelmanetal90,giovanonikazanas90,atoyandermer03},
and they are unaffected by adiabatic energy losses that the protons may sustain
in the system as it expands \citep{rachenmeszaros98}. For the purposes of this
paper, we will assume that ``neutron conversion'' is the injection mechanism of
CRs from BL Lacs into the intergalactic medium 
\citep[e.g.][]{kis14}, while we neglect any additional contribution to the
neutrino and CR fluxes from direct proton escape
\citep[e.g.][]{Essey_2010,Kalashev_2013}.  This is a valid assumption as long as
the escaping protons are susceptible to adiabatic energy losses (e.g. the
emitting region lies within an expanding jet) \citep{rachenmeszaros98} and may
end up carrying a negligible fraction of the UHECR flux. In this regard, the CR
flux estimates that follow should be considered as a lower limit.

The neutron spectrum, and thus, the escaping CR spectrum from each BL Lac,
can be directly related to the neutrino spectrum \cite[see e.g. Fig. 9
  in][]{dimi12}, at least in the optically thin regime for $\pg$
interactions\footnote{The BL Lac emitting region is optically thin to $\pg$
  interactions with typical optical depths $\tau_{\pg} \sim 10^{-6}-10^{-4}$
  \citep[see e.g.][]{atoyandermer01, Petro_2015}.}.  Under certain
simplifying assumptions, namely: (1) $E_{\rm n} \simeq 20 E_{\nu}$; (2)
$E_{\rm p} \simeq E_{\rm n}$; and (3) production of one $\pi^{\pm}$ pair per
interaction, which leads to $E_{\rm n}dN/dE_{\rm n} \approx (1/6)
E_{\nu}dN/dE_{\nu}$ \citep[see also][]{kis14}, we may write \eqb E_{\rm p}^2
\frac{dN}{dE_{\rm p}} \approx \frac{20}{6}E_{\nu}^2\frac{dN}{dE_{\nu}}.
\label{relation}
\eqe

Having already calculated the NBG flux in Sect.~\ref{sec:NBG}, we can easily
apply the above relation to all BL Lacs and estimate their contribution to the
injected CR spectrum without taking into account any propagation effects. This
will be a ``copy'' of the total NBG spectrum shown in Figs. \ref{fig:background}
and \ref{fig:background_2}, translated to higher energies by a factor of $\sim
20$ and scaled in flux by a factor\footnote{Equation \ref{relation} was derived
  taking into account neutrinos of all-flavours, whereas the NBG flux shown in
  Figs. 1 and 2 is per neutrino flavour.} of $10$.

The resulting CR spectrum from HBL for our benchmark case peaks at $E_{\rm p}
\sim 2.5\times 10^{18}$~eV with a flux $\sim 3.0\times 10^{24}$~eV$^2$ m$^{-2}$
s$^{-1}$ sr$^{-1}$, only slightly above the observational data
\citep[see][]{HIRES,Tel_Array,Auger2013}. 
A  proper comparison would require 
the calculation of the expected proton fluxes at Earth taking into account propagation effects.
Namely: 
(i) photopion \citep[e.g.][]{mueckeetal99} and photopair \citep{Blumenthal_1970} 
production on the background photon fields (cosmic microwave, cosmic infrared, and 
cosmic optical backgrounds); 
(ii) adiabatic energy losses; and (iii) magnetic deflections
due to the intergalactic magnetic field.

From the above, it becomes evident that a detailed treatment of LBL is
also required, since the specifics of their SED modelling (e.g. the maximum
proton energy) will affect our predictions of the UHECR spectrum.  We plan to
calculate the CR flux predicted by our scenario in detail in a future
paper.

\subsection{Model predictions}\label{sec:predictions}

Our model makes specific predictions on the detectability of $\en > 2$ PeV
events, that is above the current maximum energy of IceCube neutrinos. 
We have used the effective areas from \cite{ICECube13} and our
NBG to predict the number of $2 < \en < 10$ PeV neutrino events IceCube should
see (as 10 PeV is the maximum energy for which IceCube effective areas are
available).  We get $N_{\nu} \sim 4.6$ events for our benchmark case assuming
the Glashow resonance \citep{glashow_1960} is not relevant since we are dealing
with proton-photon interaction \citep{Anchordoqui_2005} (and $N_{\nu} \sim 7$
otherwise). For our other case ($Y_{\nu\gamma} = 0.8$ and $E_{\rm break} = 100$
GeV, $\Delta \Gamma = 1.0$) we derive $N_{\nu} \sim 4.0$ (or 6.1) events. Since
the NBG of our models peak at $\en > 10$ PeV by making an educated guess on the
effective areas we estimate an additional $2 - 3$ events up to $\sim 100$
PeV. Noting that the 3$\sigma$ upper limit for 0 events is 6.6 \citep{geh86}, we
are actually close to being inconsistent with the IceCube
non-detections. However, our number is probably biased because $Y_{\nu\gamma} =
0.8$ is likely an upper limit. This was derived, in fact, from a small sample of
BL Lacs, which might represent the tip of the iceberg in terms of neutrino
emission, as they were selected as the most probable candidates. For example, if
$Y_{\nu\gamma} = 0.3$ then for our benchmark case $N_{\nu} \approx 3$ (4) for $2
< \en < 100$ PeV, which is well within the 2$\sigma$ limit for 0 events. In any
case, our predictions should easily be testable by IceCube in the next few years.
  
Turning things around, we showed that the shape and the peak energy of the
neutrino spectrum expected from a single BL Lac are mostly determined by its
low-energy emission and the Doppler factor of the source (see
Sect. \ref{sec:hadro}).  Thus, the calculation of the diffuse neutrino emission
from BL Lacs can be considered to be robust. Most importantly though, our method
has only one tuneable parameter, namely the ratio $Y_{\nu\gamma}$
(eq. \ref{eq:flux}). A comparison of the model predicted NBG with current
IceCube upper limits and, ultimately, future detections at $E_{\nu} > 2$~PeV,
can be used to constrain the value of $Y_{\nu \gamma}$. In other words, this
would provide an indirect way of probing the origin of the BL~Lac $\gamma$-ray
emission.
  
\subsection{The big picture}

\begin{figure}
\hspace{-1.5cm}
\includegraphics[height=7.6cm]{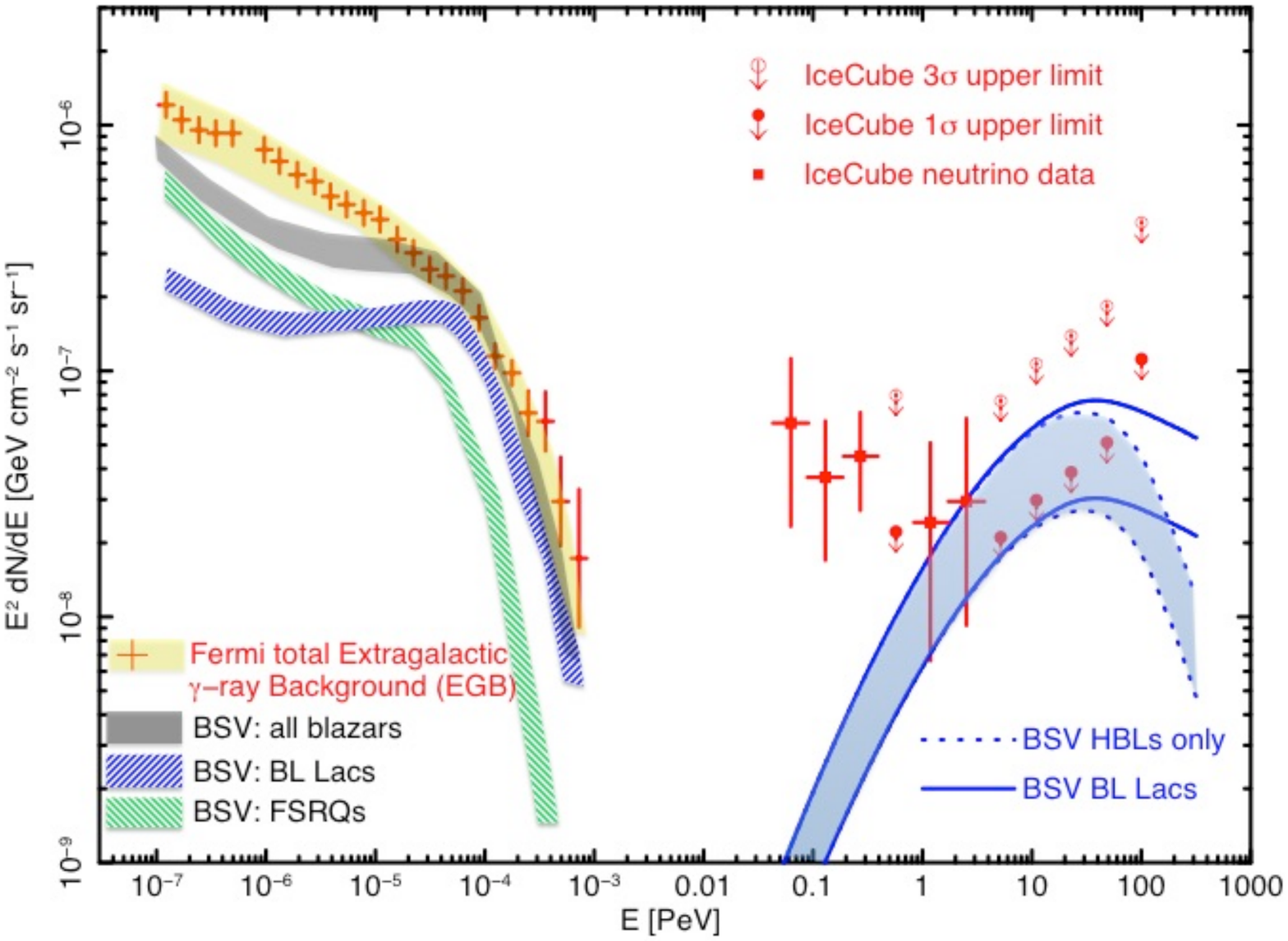}
\caption{The electromagnetic and neutrino extragalactic backgrounds predicted by
  our simulations in the energy range 100 MeV -- 300 PeV. The left side of the
  plot ($E < 1$~TeV) shows the $\gamma$-ray background for various classes
  compared to the total extragalactic electromagnetic emission observed by
  \fermi-LAT (adapted from \paperfour), whereas the right side ($E>10$~TeV)
  illustrates our prediction for the neutrino background (all flavours) for our
  benchmark case ($E_{\rm break} = 200$ GeV, $\Delta \Gamma = 0.5$) for all BL
  Lacs (blue solid line) and HBL (blue dotted line) and $Y_{\nu\gamma}$ ranging
  between 0.8 (upper curves) and 0.3 (lower curves; see text for details).
The (red) filled points are the (all flavours) data points from \protect\cite{ICECube14}, while
the open points are the $3\sigma$ upper limits.}
\label{fig:background_5}
\end{figure}

Fig.~\ref{fig:background_5} displays both the electromagnetic and neutrino
extragalactic backgrounds predicted by our simulations and the available measurements 
in the energy range 100
MeV -- 300 PeV. The left side shows the $\gamma$-ray background compared to the
total extragalactic electromagnetic emission observed by \fermi-LAT (adapted
from \paperfour), whereas the right side illustrates our prediction for the NBG
(all flavours) for our benchmark case for all BL Lacs (blue solid line) and HBL
(blue dotted line) and $Y_{\nu\gamma}$ ranging between 0.8 (upper curves) and
0.3 (lower curves), the latter value being more consistent with the IceCube
high-energy non-detections.

The EGB can be approximated by a power law with exponential cutoff having
$\Gamma \simeq 2.3$ and a break energy $\sim 280$ GeV \citep{fermiegb}, the
latter very likely due to the EBL absorption of $\gamma$-ray photons from
distant ($z \ga 0.3$) sources \citep[e.g.][and Paper IV]{ajello2015}. As a
simple extrapolation of the EGB power law to the PeV energy range goes through
the IceCube data, it might be tempting to assume that there is a single class of
sources that explains both the EGB at $E \la 10$~GeV and the NBG below 
$\sim 0.5$ PeV. This population cannot be the blazar one, for the following two
reasons: (i) in the BSV scenario, blazars contribute $\sim 50\%-70\%$ to the
total EGB at $E \la10$~GeV, while BL Lacs may explain almost $100\%$ of the EGB
flux at $E \ga 100$ GeV; (ii) similarly, BL Lacs contribute only $\sim 10\%$ to
the NBG at energies $< 0.5$~PeV, while they may fully explain the observed NBG
above 0.5 PeV. If starburst galaxies can explain part of the EGB at $E > 100$
MeV \citep[e.g.][and references therein]{lacki_2014} then they could also be a
promising candidate class for explaining the sub-PeV IceCube neutrinos
\citep[e.g.][]{loeb_2006,stecker_2007}. We note, in fact, that in proton-proton
scenarios of $\gamma$-ray emission, relevant to starburst galaxies, the neutrino
and $\gamma$-ray spectra have the same power law index as the parent proton
population \citep[e.g.][]{kelner_2006}.

Alternatively (or at the same time) the low-energy neutrino events could also
have a Galactic component \citep[e.g.][]{Pad_2014}. In any case, if there is a
different class of sources contributing to the sub-PeV energy range, there is
still room for individual BL Lac sources, like MKN 421 (see
Fig.~\ref{fig:background_2} and relevant discussion). Finally, we note that the
EGB in Fig.~\ref{fig:background_5} shows the sum of unresolved and resolved
$\gamma$-ray emission of the extragalactic sky, whereas in the case of IceCube
neutrinos we are not yet in the position to distinguish between a resolved and
an unresolved contribution. The current status of neutrino astronomy, therefore, somewhat
resembles that of $\gamma$-ray astronomy in its very early days (i.e. those
of SAS-2 and COS-B).
 
The scenario, which appears to emerge by comparing our model NBG with the data
is the following: at low energy ($\en \la 0.5$ PeV) BL Lacs can only explain
$\sim 10\%$ of the IceCube data. Some other population/physical mechanism needs
to provide the bulk of the neutrinos. However, this does not exclude the
possibility that individual BL Lacs still make a contribution at the $\approx
20\%$ level to the IceCube events.
At high energy ($\en \ga 0.5$ PeV) BL Lacs can account fully for the IceCube
data. The strong implications of our scenario are: 1. IceCube should soon start resolving
at least some of the NBG; 2. IceCube should also detect events at $\en \ga 2$ PeV 
in the next few years.

\section{Conclusions}

We have included in the {\it blazar simplified view} scenario, which reproduces
extremely well the statistical properties of blazars from the radio to the
$\gamma$-ray band, a hadronic component and calculated via a leptohadronic model
the neutrino background produced by the whole BL Lac class. For the first time,
this is done by summing up the fluxes of all the BL Lacs generated by a Monte
Carlo simulation, each with its own different properties. Our main results can
be summarised as follows:

\begin{enumerate} 
\renewcommand{\theenumi}{(\arabic{enumi})} 
 
\item BL Lacs as a class can easily explain the whole neutrino background at
  high-energies ($\ga 0.5$ PeV) while they do not contribute much ($\sim 10\%$)
  at low-energies ($\la 0.5$ PeV).

\item Individual BL Lacs, however, including some of the sources selected as
  possibly associated with some IceCube events by \cite{Pad_2014}, can still
  make a contribution at the $\approx 20\%$ level to the low-energy events.

\item Given our Monte Carlo approach, we can easily characterise the BL Lacs
  producing most of the neutrino background. These are of the HBL type (up to
  $\en \approx 30$ PeV), have relatively low redshifts ($\langle z \rangle \sim
  1.3$), and about half of them have their optical spectra swamped by
  non-thermal emission and therefore an unmeasurable redshift. The (small)
  fraction of {\it Fermi} detectable sources in our scenario is consistent with
  a preliminary IceCube likelihood analysis on {\it Fermi} blazars.

\item A simultaneous look at the $\gamma$-ray and neutrino backgrounds leads us
  to suggest that another population/physical mechanism could explain both the
  former at $E \la 10$ GeV (since blazars dominate above that energy) and the
  latter at $\en \la 0.5$ GeV. This might include star forming galaxies,
  although a Galactic component for the low-energy IceCube events could also be
  possible.

\item Our simulations at face value predict that IceCube should soon start
  resolving a fraction of the neutrino background and detect events at $\en > 2$
  PeV. A non-detection will constrain the value of our only tuneable parameter
  ($Y_{\nu\gamma}$) and will provide an indirect way of probing the origin of
  $\gamma$-ray emission in BL~Lacs.

\end{enumerate}

\section*{Acknowledgments}

We thank Stefan Coenders, Gabriele Ghisellini, Tullia Sbarrato, and an
  anonymous referee, for useful comments. Support for this work was provided
by NASA to MP through Einstein Postdoctoral Fellowship grant number
PF3~140113 awarded by the Chandra X-ray Center, which is operated by the
Smithsonian Astrophysical Observatory for NASA under contract NAS8-03060. ER
is supported by a Heisenberg Professorship of the Deutsche
Forschungsgemeinschaft (DFG RE 2262/4-1).

\label{lastpage}
\end{document}